\documentclass[preprint,twocolumn,pre,10pt]{revtex4}%
\usepackage{graphicx}
\usepackage{dcolumn}
\usepackage{bm}
\usepackage{amsmath}
\usepackage{amsfonts}
\usepackage{amssymb}%
\begin{document}
%
\preprint{Marc Durand}
\title
[Optimizing transport in a homogeneous network]{Optimizing transport in a homogeneous network}%
%

\author{Marc Durand}%
%

\email{mdurand@deas.harvard.edu}%
%

\affiliation{DEAS, Harvard University, 29 Oxford St., 02138 Cambridge MA, USA}%
%

\author{Denis Weaire}%
%

\affiliation{Trinity College Dublin, Department of Physics, Dublin 2}%
%

%

%

\begin{abstract}%
Many situations in physics, biology, and engineering consist of the transport
of some physical quantity through a network of narrow channels. The ability of
a network to transport such a quantity in every direction can be described by
the \textit{average conductivity }associated with. When the flow through each
channel is conserved and derives from a potential function, we show that there
exist an upper bound of the average conductivity, and explicitly give the
expression for this upper bound as a function of the channel permeability and
channel length distributions. Moreover, we express the necessary and
sufficient conditions on the network structure to maximize the average
conductivity. These conditions are found to be independent of the connectivity
of the vertices.%
\end{abstract}%
%

\volumeyear{year}%
%

\volumenumber{number}%
%

\issuenumber{number}%
%

\eid{identifier}%
%

\date[Dated: ]{April 12, 2004}%
%

\received[Received text: ]{April 12, 2004}%
%

\revised[Revised text]{date}%
%

\accepted[Accepted text]{date}%
%

\published[Published text]{date}%
%

\startpage{1}%
%

\endpage{4}%
%

\maketitle

Examples of transport phenomena through a network of channels abound in nature
and engineering: blood flow through the microcirculatory system, water
transport through the venation of a leaf, water and electricity supply in a
city, heat conduction through an open cell material, etc. Therefore, the
search for a network structure optimizing the transport processes may be
useful for a better understanding of the structure of natural networks and for
the conception of optimized materials. Different functions can be optimized,
such as dissipated power, volume or surface area of the channels. Different
constraints can be imposed: topology, flow rate, volume, etc. Various models
have been proposed to understand the structure of natural networks\cite{Zamir,
Changizi}, mostly based on the assumption of \textit{local} optimization. The
idea is to move the position of a given junction, with the other junctions and
the topology fixed, in order to optimize one of the above functions. In a
previous work, we considered the \textit{global} optimization of the whole
network structure for its electrical property \cite{Durand}. We showed the
existence of an upper bound for the average electrical conductivity of a
network made of uniform wires, for a given amount of material per unit volume,
and proved two necessary and sufficient conditions on the structure of the
network for the upper bound to be reached. We then used these results to
derive the expression of electrical conductivity of dry foams. In the present
paper, we generalize the idea of an optimal structure for transport processes
to networks made of non-uniform pipes and to any flow process, and derive
three more general necessary and sufficient conditions for maximizing the
transport properties. Surprisingly, we find that these conditions do not
depend on the connectivity of the junctions.

Let us first denote each pipe by a pair of indices $(i,j)$ corresponding to
the labels of its two ends. We consider pipes for which the aspect ratio is
very large, so we can univocally define for each pipe $(i,j)$ a length
$l_{ij},$ a local cross-sectional area $s_{ij}(l)$, and a \textit{local
permeability} $k_{ij}\left(  l\right)  $ ($k_{ij}\left(  l\right)  $ is
function of $s_{ij}\left(  l\right)  $ and both can vary with the curvilinear
coordinate $l$ along the channel). Let us assume that the flow through each
channel is directly related to the gradient of a potential function $V$, so
the \textit{flow }vector $\mathbf{I}$, the permeability and $V$ are related
by:
\begin{equation}
\mathbf{I}=-k_{ij}(l)\mathbf{\nabla}V.
\end{equation}
Moreover we suppose that the flow is in a steady state (so $V$ satisfies
Laplace equation: $\mathbf{\nabla}^{2}V=0$), and is conserved through each
pipe and each junction. We then define the ''\textit{dissipated power}''
associated with the flow in the network as:
\begin{equation}
P=-\sum\limits_{(i,j)}\int_{0}^{l_{ij}}\mathbf{I\cdot\nabla}Vdl.
\end{equation}
Using the assumption of steady state flow, we can rewrite the dissipated power
as:
\begin{equation}
P=\sum\limits_{(i,j)}u_{ij}i_{ij},
\end{equation}
where $u_{ij}$ and $i_{ij}$ are respectively the potential difference and the
flow rate through the channel $(i,j)$. Since the flow is conserved through
each pipe and each junction, the total dissipated power in the network is
equal to the product $U.I$, where $U$ is the potential difference between the
inlet (source) and the outlet (sink) of the network and $I$ is the crossing
flow rate, as for the dissipated power of an electrical network. Pursuing the
analogy, it follows that the actual distribution of flow rates for a given
total flow rate is the one which minimize the dissipated power, and
subsequently the monotonicity law of Rayleigh is valid
\cite{Rayleigh,Jeans,Durand}: \textit{If any of the resistances of a circuit
are increased, the effective resistance between any two points can only
increase. If they are decreased, the effective resistance can only decrease}.
Although the monotonicity law cannot predict the sign of the effective
resistance variation when some resistances are increased and some others are
decreased (the variation obviously depends on the structure of the network),
it will be helpful to determine the optimal geometry of a network for its
transport properties.

\subsection{Maximal average conductivity of a network}

\begin{figure}
[h]
\begin{center}
\includegraphics[
height=1.8144in,
width=3.3797in
]%
{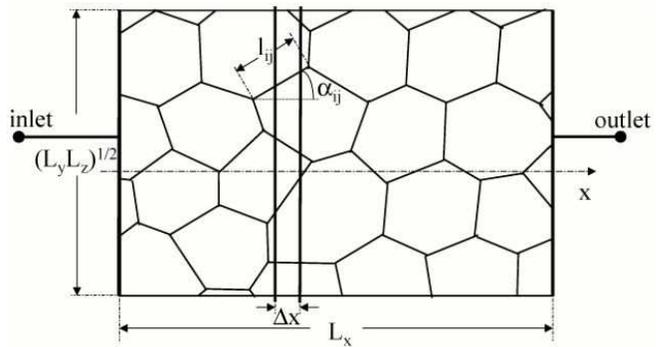}%
\caption{Schematic representation of a network shorted with parallel sheets of
infinite conductivity. The resistance of the truncated channel $(i,j)$ is
equal to $\Delta x/\left(  k_{ij}(l)\cos\alpha_{ij}\right)  $.}%
\label{Fig1}%
\end{center}
\end{figure}

On a macroscopic scale, the network can be seen as a continuous medium,
\textit{a priori} anisotropic, whose transport properties are described by an
effective conductivity tensor $\underline{\underline{\sigma}}$ - the
conductivity being defined as the permeability per unit length (2D network) or
per unit area (3D network). The \textit{average conductivity} associated with
this tensor is defined as:
\begin{equation}
\sigma_{m}=\left\langle \mathbf{u}.\underline{\underline{\sigma}}%
.\mathbf{u}\right\rangle ,
\end{equation}
where the brackets indicate that the term inside is averaged in all the
directions of the unit vector $\mathbf{u}$. This parameter can be simply
related to the trace of $\underline{\underline{\sigma}}$ and the dimension of
space $d$ \cite{endnote} by:
\begin{equation}
\sigma_{m}=\frac{1}{d}Tr\left[  \underline{\underline{\sigma}}\right]  .
\end{equation}
The average conductivity characterizes the ability of the network to transport
the physical quantity associated with in all the directions. We shall show the
existence of an upper bound for $\sigma_{m}$, which can be expressed as a
function of the channel permeability and channel length distributions:
\begin{equation}
\sigma_{m}\leq\frac{1}{d}\sum\limits_{(i,j)}\frac{l_{ij}\overline{k}_{ij}%
}{\emph{L}^{d}}. \label{upper bound}%
\end{equation}
$\emph{L}^{d}$ is the hypervolume of the network on which the conductivity
tensor is defined, and $\overline{k}_{ij}$ is the average permeability of the
channel $\left(  i,j\right)  $ defined by:
\begin{equation}
\overline{k}_{ij}=\frac{1}{l_{ij}}\int_{0}^{l_{ij}}k_{ij}(l)dl.
\end{equation}
We present here the demonstration for a three-dimensional network ($d=3$). We
first study the case of a network made of straight pipes: imagine that this
network is shorted with thin parallel sheets of infinite conductivity,
perpendicular to the direction $x$ of the applied potential difference, and
separated from each other by $\Delta x$, as in Fig.\ref{Fig1} ($\Delta x$ has
to be small compared to the typical channel length, but large compared to the
typical channel diameter, so the flow lines in a given channel are mostly
along its axis).%
According to the monotonicity law, the conductivity of this shorted network is
higher than the conductivity of the original one. Furthermore the potential is
uniform on each sheet and the resistance $\Delta R(x)$ of the network slice at
position $x$ corresponds to the parallel association of the truncated channels
that it contains. Since the resistance corresponding to a given truncated
channel $(i,j)$ is equal to $\Delta x/\left(  k_{ij}(l)\cos\alpha_{ij}\right)
$, where $\alpha_{ij}$ is the angle between the channel $(i,j)$ and the axis
$x$ (the angle is defined such that $\cos\alpha_{ij}\geq0$), we have:
\begin{equation}
\frac{1}{\Delta R(x)}=\sum_{(i,j)}P(x;x_{i},x_{j})\frac{k_{ij}(l)\cos
\alpha_{ij}}{\Delta x}, \label{delta R}%
\end{equation}
where the sum is carried out on all the channels of the network by introducing
the function $P(x;x_{i},x_{j})$ which take the value $1$ if the channel
$(i,j)$ is intersected by the equipotential plane passing by $x$ (i.e. if $x$
is between $x_{i}$ and $x_{j}$) and $0$ otherwise. The total resistance is
given by the sum of the slice resistances. Using the fact that the product of
the average of a set of positive values\noindent\noindent\ $\{f_{1}%
,f_{2},...,f_{N}\}$ by the average of their inverses is always greater or
equal to one:%

\begin{equation}
\left(  \frac{1}{N}\sum_{k=1}^{N}f_{k}\right)  \left(  \frac{1}{N}\sum
_{k=1}^{N}\frac{1}{f_{k}}\right)  \geq1, \label{inequality}%
\end{equation}
and passing to the continuum limit, we obtain a lower bound for the resistance
of the shorted network:
\begin{equation}
\frac{1}{R_{x}}\leq\int_{0}^{\infty}\sum_{(i,j)}P(x,x_{i},x_{j})k_{ij}%
(l)\cos\alpha_{ij}\frac{dx}{L_{x}^{2}}.
\end{equation}
We can switch the sum and the integral of this expression, and it follows
after integration that:
\begin{equation}
\sigma_{x}^{(s)}\leq\frac{1}{\emph{L}^{3}}\sum_{(i,j)}\overline{k}_{ij}%
l_{ij}\cos^{2}\alpha_{ij}, \label{shorted x}%
\end{equation}
where the conductivity of the shorted network in the $x$ direction is defined
by $\sigma_{x}^{(s)}=\frac{L_{x}}{L_{y}L_{z}R_{x}}$, $L_{y}$ and $L_{z}$ being
the network lengths in directions $y$ and $z$, and $\emph{L}^{3}=L_{x}%
L_{y}L_{z}$ the volume on which the conductivity is defined. Using the same
arguments in the two other directions and the fact that the sum of the squared
direct cosines is equal to one, we see that inequality (\ref{upper bound}) is
true for a network made of straight channels. It is clear that the average
conductivity of a network with curved channels is bounded too: in this case,
we can built a new network by keeping the junctions fixed in their positions
but linked with straight channels. Let $l_{ij}^{\prime}$ be the length of the
straight channel $(i,j)$. Moreover we can choose its local permeability
$k_{ij}^{\prime}(l)$ to be equal to the local permeability of the
corresponding curved channel on an arbitrary portion $l_{ij}^{\prime}$ of its
length, say: $k_{ij}^{\prime}(l)=k_{ij}(l)$ for $0\leq l\leq l_{ij}^{\prime}$.
So the resistance of the straight channel, equal to $\int_{0}^{l_{ij}^{\prime
}}\frac{dl}{k_{ij}^{\prime}(l)}$, is lower than the original curved channel.
We know the conductivity $\sigma_{m}^{\prime}$ of such a network is bounded:
\begin{equation}
\sigma_{m}^{\prime}\leq\frac{1}{3}\sum\limits_{(i,j)}\frac{l_{ij}^{\prime
}\overline{k^{\prime}}_{ij}}{\emph{L}^{3}}.
\end{equation}
On the one hand the conductivity of this network is higher than the
conductivity $\sigma_{m}$ of the original network (from the monotonicity law),
and on the other hand $l_{ij}^{\prime}\overline{k^{\prime}}_{ij}\leq
l_{ij}\overline{k}_{ij}$, so inequality (\ref{upper bound}) holds for a
network made of curved channels as well.

\subsection{Optimizing transport}

In addition to the existence and the expression of an upper bound for the
average conductivity, we show that the average conductivity reaches the upper
bound if and only if the following three conditions are satisfied:

\begin{enumerate}
\item[a)] Each channel has a uniform local permeability along the channel:
$k_{ij}(l)=\overline{k}_{ij}=k_{ij}$

\item[b)] All the channels are straight.

\item[c)] Every junction $(i)$ between channels satisfies $\sum_{j}%
k_{ij}\mathbf{e}_{ij}=\mathbf{0}$, where $\mathbf{e}_{ij}$ are
outward-pointing unit vectors in the directions of adjoining channels.
\end{enumerate}

Note that the last condition is equivalent to a force balance equation at each
vertex, the weight of the force pulling along a channel being analogous to its
permeability. Furthermore, it is worth noting that the three conditions are
independent of the connectivity of the junctions. As an illustrative example
of this property, the periodic square, hexagonal and triangular networks built
with a same set of channels have the same conductivity.

\subparagraph{Necessity of the conditions:}

Let us suppose that the conductivity of the network is equal to its maximal
value:
\begin{equation}
\sigma_{m}=\frac{1}{d}\sum\limits_{(i,j)}\frac{l_{ij}\overline{k}_{ij}%
}{\emph{L}^{d}},
\end{equation}
and let us make some infinitesimal changes in the network structure. First,
suppose we vary the local permeability of a given channel $(i,j)$ by a small
amount $\delta k_{ij}(l)$, without altering its length $l_{ij}$ or its average
permeability $\overline{k}_{ij}$, so:
\begin{equation}
\int_{0}^{l_{ij}}\delta k_{ij}(l)dl=0.
\end{equation}
Since $l_{ij}$ and $\overline{k}_{ij}$ remain constant, the corresponding
variation $\delta\sigma_{m}$ of the conductivity is zero. But obviously we can
choose a variation $\delta k_{ij}(l)$ such that the resistance of the channel
is increased:
\begin{equation}
\int_{0}^{l_{ij}}\delta\left(  \frac{1}{k_{ij}(l)}\right)  dl\geq0.
\end{equation}
In order for the monotonicity law to be satisfied, the variation of the
resistance has to be zero, which can be mathematically expressed as:
\begin{equation}
\int_{0}^{l_{ij}}\left(  -\frac{1}{k_{ij}^{2}(l)}+\lambda\right)  \delta
k_{ij}(l)dl=0,
\end{equation}
where $\lambda$ is a Lagrange multiplier. Since this equality has to be true
for any variation $\delta k_{ij}(l)$, it follows that $k_{ij}(l)$ has to be
uniform along the channel, and so condition a. is indeed required.

Necessity of conditions b) and c) is proved by means of arguments similar to
those used in our previous work\cite{Durand}: first, imagine that we change
the length of a given channel $(i,j)$ of a network for which condition a) is
satisfied, the positions of all junctions and the lengths of all the other
channels remaining unaltered. To this variation of length $\delta l_{ij}$
corresponds a variation of the conductivity:
\begin{equation}
\delta\sigma_{m}=\frac{1}{\emph{L}^{d}d}k_{ij}\delta l_{ij},
\end{equation}
implying that $\delta\sigma_{m}$ and $\delta l_{ij}$ have the same sign.
However, if the length of the channel is increased ($\delta l_{ij}\geq0$), the
resistance of the channel is increased too, and it follows from the
monotonicity law that the conductivity can only decrease ($\delta\sigma
_{m}\leq0$). So the variation of the length has be zero at first order,
implying the necessity of condition b).

Now, imagine a network for which conditions a) and b) are satisfied. In the
previous section, we showed the existence of an upper bound of the
conductivity in the direction $x$ for a network made of straight channels, by
using two successive inequalities: first, the conductivity\ of such a network
is lower than the conductivity of the same network intersected with zero
resistance sheets; second, the shorted network conductivity itself is bounded,
using the fact that the equivalent resistance of $N$ resistive elements in
series arrangement is greater or equal to the equivalent resistance of the
same resistive elements in parallel arrangement times $N^{2}$ (inequality
(\ref{inequality})). So in order to get the exact upper bound, these two
inequalities have to become strict equalities. The first one implies that the
presence of sheets does not modify the distributions of potentials, and so the
potential in the channels is a function of $x$ only. To see this, increase
progressively the resistance of the sheets up to infinity (what corresponds to
the initial network). From the monotonicity law, this can only decrease the
conductivity of the network. The only way for the conductivity to stay at its
maximum value when increasing the resistance of each sheet is by having no
current through them, and so the potential in the channels is a function of
$x$ only. The second equality requires that the resistance of every slice (of
equal thickness) is the same, or equivalently, the resistance of a slice of
arbitrary thickness $x$ is simply proportional to $x$. Hence the potential is
indeed \textit{linear} in $x$. Examination of flow conservation at a vertex in
such a potential immediately leads to condition :
\begin{equation}
\sum_{j}k_{ij}\cos\alpha_{ij}.sgn\left(  x_{i}-x_{j}\right)  =0, \label{eqx}%
\end{equation}
(the term $sgn\left(  x_{i}-x_{j}\right)  $ is introduced in order to satisfy
$\cos\alpha_{ij}\geq0$). This equality is nothing but the projection of
condition c) on the axis $x$. Since the same argument holds in the two other
directions, we prove that condition c) is indeed required.%
\begin{figure}
[h]
\begin{center}
\includegraphics[
height=2.2701in,
width=3.3702in
]%
{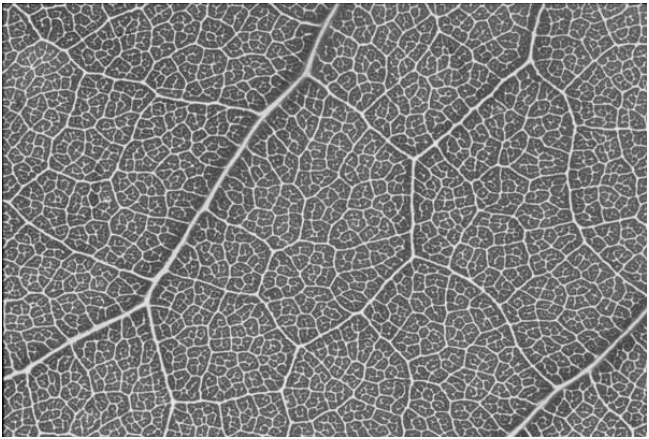}%
\caption{Netted venation of a leaf. The angles between adjacent veins are
correlated to their cross-section areas \cite{Bohn}.}%
\label{leaf venation}%
\end{center}
\end{figure}

\subparagraph{Sufficiency of the conditions:}

Now consider a network for which conditions a), b), c) are satisfied, and
suppose that a potential difference $U_{x}$ is applied between the two
regarding faces orthogonal to $x$. Let us show that the trial potential
function defined as $\phi=-\frac{U_{x}}{L_{x}}x$ is the physical solution. We
first check that the flow is conserved at each junction under the stated
conditions: the current in the straight channel $(i,j)$ is given by:
\begin{equation}
I_{ij}=-k_{ij}\mathbf{\nabla}\phi\cdot\mathbf{e}_{ij}=k_{ij}\frac{U_{x}}%
{L_{x}}\mathbf{e}_{x}\cdot\mathbf{e}_{ij},
\end{equation}
so:
\begin{equation}
\sum_{j}I_{ij}=\frac{U_{x}}{L_{x}}\mathbf{e}_{x}\cdot\sum_{j}k_{ij}%
\mathbf{e}_{ij}=0.
\end{equation}
The trial potential function also satisfies the boundary conditions, and so is
the correct physical solution. Now we can show that the average conductivity
is equal to the upper bound. The potential is uniform on planes perpendicular
to the direction $x$, and so the system is unaltered when intersected by thin
parallel sheets of infinite conductivity orthogonal to that direction. We
previously calculated the elementary resistance\ $\Delta R(x)$ of a slice of
thickness $\Delta x$ at position $x$ for such a network (Eq. \ref{delta R}).
The global flow rate $I_{x}$, the elementary resistance $\Delta R(x)$, and the
elementary potential difference $\Delta\phi$\ across the slice are related
by:
\begin{equation}
\frac{\Delta\phi}{\Delta x}=\frac{\Delta R}{\Delta x}I_{x}.
\end{equation}
But here both the potential gradient and the global current are independent of
$x$, and so is $\frac{\Delta R}{\Delta x}$. Using Eq. (\ref{delta R}) and
integrating in the $x$ direction, we finally get the expression of the
conductivity in the $x$ direction:
\begin{equation}
\sigma_{x}=\frac{1}{\emph{L}^{d}}\sum_{(i,j)}k_{ij}l_{ij}\cos^{2}\alpha_{ij}.
\end{equation}
This expression corresponds indeed to the upper bound of the conductivity
along $x$, as expressed in (\ref{shorted x}). The same reasoning can be
applied for the conductivity in the two others directions, and so the
sufficiency of conditions a), b) and c) is proven.

Let us apply the preceding results to two very common flow profiles, namely
plug flow and Poiseuille flow. Many transport phenomena through pipes are
described by plug flow, such as the flow of fluids in porous conducts, the
flow of heat by conduction, the electrical current, or diffusive flow. All
these flows are engendered by a gradient of a potential function (such as
pressure, temperature, electric potential, material concentration), and the
permeability for all these situations is directly proportional to the
cross-sectional area. Moreover, if the conductivity has the same value
$\sigma_{0}$ for all pipes (so $k_{ij}=\sigma_{0}s_{ij}$), the upper bound of
the average conductivity simplifies to
\begin{equation}
\sigma_{m}\leq\frac{1}{d}\sigma_{0}\varepsilon,
\end{equation}
where $\varepsilon$ is the volume fraction of the continuous phase. As a
consequence, the maximal conductivity does not depend on the permeability and
length channel distributions, but only on the total amount of material that
the network contains (there exists a universal upper bound of the quantity
$\left(  \sigma_{m}/\varepsilon\right)  $, the average conductivity per unit
volume of material). Note that the expression of the upper bound is equivalent
to the Hashin-Shtrikman bound for the electrical conductivity of an
heterogenous material in the limit of small volume fraction of the conducting
phase\cite{Hashin, Torquato}.

Condition c) leads to different optimal structures depending on the kind of
flow profile in the channels. In the case of a plug flow profile with idential
conductivity $\sigma_{0}$, this condition becomes:
\begin{equation}
\sum_{j}s_{ij}\mathbf{e}_{ij}=\mathbf{0}. \label{cond plug}%
\end{equation}
In the case of Poiseuille flow, the permeability varies like the square of the
cross-sectional area. Thus, if we assume again the same conductivity\ $\sigma
_{0}$\ for every pipe, condition c) becomes:
\begin{equation}
\sum_{j}s_{ij}^{2}\mathbf{e}_{ij}=\mathbf{0}. \label{cond pois}%
\end{equation}

As a concluding remark, it is worth noting that some natural networks, like
leaf venation\cite{Bohn,Couder}, have a very well defined structure, in the
sense that the angles between adjacent veins are correlated to their
cross-sectional areas (see Fig. \ref{leaf venation}). This fact presumably
corresponds to some optimization principle. The conditions for transport
optimization presented here may give an explanation for these typical
patterns. In the particular case of leaf venation, veins are composed of a
complex tangle of smaller tubes \cite{Xylems}, hence we may expect the
relation (\ref{cond plug}) rather than (\ref{cond pois}) to be satisfied
(although veins are not fully impermeable). Alternatively, this correlation
between angles and cross-section areas could be explained by the optimization
of the mechanical stability of the leaf \cite{Roth,Durand2}. Experimental
study of the leaf venation structure is subject to a current investigation to
be compared with the transport optimization criteria presented in this paper.

M.D. thanks H. A. Stone for encouraging this research.


\begin{thebibliography}{99}                                                                                               %
\bibitem {Zamir}M. Zamir, \textit{J. theor. Biol.} \textbf{62}, 227-251 (1976).

\bibitem {Changizi}M. A. Changizi and C. Cherniak, \textit{Can. J. Physiol.
Pharmacol.} \textbf{78}, 603-611 (2000).

\bibitem {Durand}M. Durand, J.-F. Sadoc, and D. Weaire, \textit{Proc. R. Soc.
Lond. A. }\textbf{460}, 1269-1285 (2004).

\bibitem {Rayleigh}J. W. S. Rayleigh, On the theory of resonance. In
\textit{Collected scientific papers}, volume 1, 33-75 (1899).

\bibitem {Jeans}J. H. Jeans, \textit{The mathematical theory of electricity
and magnetism}, 5th edition, Cambridge University Press (1925).

\bibitem {endnote}\textit{A priori} $\sigma_{m}$ can be a function of the
trace and the determinant of $\underline{\underline{\sigma}}$, which are the
two invariants of a symmetrical 2$^{nd}$ order tensor, but using the
additivity property of tensors and dimensional analysis, it follows that
$\sigma_{m}$ can only depend on the trace, in a linearly way. The prefactor
$1/d$ comes naturally by studying the case of an isotropic network.

\bibitem {Hashin}Z. Hashin and S. Shtrikman, \textit{J. of Appl. Phys.
}\textbf{35}, 3125-3131 (1962).

\bibitem {Torquato}S. Torquato, Random Heterogeneous Materials,
Springer-Verlag, New-York (2002).

\bibitem {Xylems}A. H. de Boer and V. Volkov, \textit{Plant, Cell and
Environment} \textbf{26}, 87-101 (2003).

\bibitem {Bohn}S. Bohn, B. Andreotti, S. Douady, J. Munzinger,\ and Y. Couder,
\textit{Phys. Rev. E} \textbf{65}, 061914 (2002).

\bibitem {Couder}Y. Couder, L. Pauchard, C. Allain, M. Adda-Beddia, and S.
Douady, \textit{Eur. Phys. J. B} \textbf{28}, 135-138 (2002).

\bibitem {Roth}A. Roth-Nebelsick, D. Uhl, V. Mosbrugger, and H. Kerp,
\textit{Annals of Botany} \textbf{87}, 553-566 (2001).

\bibitem {Durand2}M. Durand (unpublished).
\end{thebibliography}
\end{document}